\documentclass[a4paper,11pt]{article}

\usepackage{jheppub} 

\title{\boldmath Tachyon Model of Tsallis Holographic Dark Energy}

\author[a]{Yang Liu}

\emailAdd{xijubear2020@Outlook.com}

\abstract{In this paper we consider the correspondence between the tachyon dark energy model and the Tsallis holographic dark energy scenario in an FRW universe.  We demonstrate the Tsallis holographic description of tachyon dark energy in an FRW universe and reconstruct the potential and basic results  of  the  dynamics  of  the  scalar  field  which  describe  the  tachyon  cosmology. In a flat universe, in the tachyon model of Tsallis holographic dark energy, independently of the existence of interaction between dark energy and matter or not, $\dot{T}^2$ must always be zero. Therefore, the equation of state $\omega_D$ is always $-1$ in such a flat universe. For a non-flat universe, $\dot{T}^2$ cannot be zero so that $\omega_D \neq -1$ which cannot be used to explain the origin of the cosmological constant. $\dot{T}^2$ monotonically decreases with the increasing of $\cos(R_h/a)$ and $\cosh(R_h/a)$ for different $\delta$s. In particular, for an open universe, $\dot{T}^2$ is always larger than zero while for a closed universe, $\dot{T}^2$ is always smaller than zero which is physically invalid. In addition, we conclude that with the increasing of $\cos(R_h/a)$ and $\cosh(R_h/a)$, $\dot{T}^2$ always decreases monotonically for irrespective of the value of $b^2$.}

\begin{document} 
\maketitle
\flushbottom

\section{Introduction}
\label{sec:intro}
Observations from type Ia supernovae [1,2,3] in association with Large Scale Structure [4,5] and Cosmic Microwave Background anisotropies [6],  have provided the main evidence that the universe is experiencing an accelerated expansion. To explain the reason for this expansion of the universe, many theories have been proposed. The most widely-accepted explanation is dark energy since dark energy has a negative pressure. However, the  nature and cosmological origin of dark energy has still not been determined.\\
The most obvious candidate for dark energy is the cosmological constant [7,8] for which $p = -\rho$, but this explanation also suffers from “fine-tuning problems”. In view of this a series of alternative proposals for dark energy have been put forward. In particular, various scalar-field dark energy models, such as quintessence [9], K-essence [10], phantom [11], tachyon [12], ghost condensate [13,14], quintom [15], interacting dark energy models [16], braneworld models [17], and Chaplygin gas models [18], have been studied.\\
One attempt at accounting for the nature of dark energy is termed “holographic dark energy” which is derived from the framework of quantum gravity. The proposal is based on the  holographic principle which states that the number of degrees of freedom related to entropy scales directly  with the enclosing area of the system [19,20]. ’t Hooft [21] and Susskind [22] have shown that effective local quantum field theories significantly over-count the number of degrees of freedom since entropy scales extensively for an effective quantum field theory in a box of size L with UV cut-off $\Lambda$. To solve this problem, Cohen et.al [23] have pointed out that the total energy of a system with size $L$ should not exceed a black hole of the same size, that is to say, $L^3 \rho_{\Lambda} \leq L M^{2}_p$. Here $M_p$ denotes the Planck mass $(M^{2}_p=\frac{1}{8 \pi G})$ and $\rho_{\Lambda}$ is the quantum zero-point energy density caused by UV cutoff $\Lambda$. The largest value of $L$ is required to define the limit of this inequality. Taking this approach we can obtain the holographic dark energy density is $\rho_{\Lambda} =\frac{3c^2 M^{2}_p}{L^2} $, where the coefficient $3$ is used merely for convenience and the parameter $c$ is dimensionless. As an application of the holographic principle in cosmology, the authors of reference [24] investigated  the consequences of excluding from the system those degrees of freedom which will never be observed by the effective field theory giving rise to an IR cut-off $L$ at the future event horizon. Thus in a universe dominated by DE, the future event horizon will tend towards a constant of the order $H_0^{-1}$ , i.e. the present Hubble radius [25]. The problem of taking the apparent (Hubble) horizon - the outermost surface defined by the null rays which are instantaneously not expanding, $R_A = 1/H $- as the IR cut-off in the flat universe was discussed by Hsu [25,26]. According to Hsu’s argument, employing the Friedmann equation $\rho= 3M^2_PH^2$ where $\rho$ is the total energy density and taking $L = H^{-1}$ we will obtain $\rho_m = 3(1 - c^2)M^2_P H^2$ [25,26].\\
Recently, using Tsallis generalized entropy [27] and by considering the Hubble horizon as the IR cutoff, in agreement with the thermodynamics considerations [28,29], a new HDE model, termed “Tsallis holographic dark energy” (THDE), has been developed and studied in the standard cosmology framework [30,31]. At first glance, it appears to be  an appropriate model for the current universe in the standard cosmology framework [30,32,33]. However, like the primary HDE based on the Bekenstein entropy [34], THDE is also unstable [30,32,33]. More studies concerning the various cosmological features of Tsallis generalized statistical mechanics can be found in ref.[35]. It is also useful to note here that an interaction between the cosmos sectors which does not involve a change of sign also cannot  produce stability for this model [33].\\
The tachyon which is unstable field  has become important in string theory through its role in the Dirac-Born-Infeld (DBI) action which is used to describe the D-brane action [12,36,37]. The cosmological model based on the effective Lagrangian of tachyonic matter 
\begin{equation}\label{eq:1.1}
L(T)= -V(T) \sqrt{1-T_{,\mu}T^{,\mu}}
\end{equation}
with the potential $V(T)=\sqrt{A}$ coincides exactly with the Chaplygin gas model [38,39]. In section 2, we review the basics of Tsallis holographic dark energy scenario. In this paper, we propose  a correspondence between the tachyon dark energy model and the Tsallis holographic dark energy scenario. In section 3 and 4, we demonstrate this holographic description of tachyon dark energy and reconstruct the potential and basic results of the dynamics of the scalar field which describe the tachyon cosmology in a flat and non-flat universe, respectively. In section 5, we discuss and summarize the results.\\

\section{The Basics}
Following ref.[30], let us review the Tsallis holographic dark energy model briefly. We consider the Friedmann-Robertson-Walker (FRW) universe metric
\begin{equation}\label{eq:2.1}
ds^2 = -dt^2 + a^2(t) \{\frac{dr^2}{1-kr^2} + r^2 d\Omega^2\} 
\end{equation}
where $k$ denotes the curvature of space whereby $k=0,1,-1$ for flat, closed and open universe respectively [25].\\
The holographic energy density (HDE) in standard cosmology is defined by
\begin{equation}\label{eq:2.2}
\rho_D = \frac{3c^2}{8\pi G L^2}= \frac{3c^2 M^2_p}{L^2} 
\end{equation}
where $c$ is a dimensionless parameter and radius $L$ is defined as
\begin{equation}\label{eq:2.3}
L = a r(t) 
\end{equation}
Here, $a$ is scale factor and $r(t)$ is relevant to the future event horizon of the universe.\\
In general, $r(t)$ and $L$ can be determined by the following relations [25]:
\begin{equation}\label{eq:2.4}
\int^{r_1}_{0} \frac{dr}{\sqrt{1 - k r^2}} = \frac{1}{\sqrt{|k|}} \sin n^{-1} (\sqrt{|k|} r_1) 
\end{equation}
In particular,
\begin{equation}\label{eq:2.5}
\int^{r_1}_{0} \frac{dr}{\sqrt{1 - k r^2}} = 
    \begin{cases}
\sin^{-1} (\sqrt{|k|} r_1)/\sqrt{|k|}  & \text{$k$=1}\\
r_1 & \text{$k$=0}\\
\sinh^{-1} (\sqrt{|k|} r_1)/\sqrt{|k|} & \text{$k$=-1}
\end{cases}       
\end{equation}.\\
And one can derive that [25]
\begin{equation}\label{eq:2.6}
L = \frac{a(t) \sin n[\sqrt{|k|}R_h(t) /a(t)]}{\sqrt{|k|}},
\end{equation}
where $R_h(t)$ is the future event horizon in flat universe, i.e. [25],
\begin{equation}\label{eq:2.7}
R_h(t) = a \int^{\infty}_{0} \frac{dt}{a} = a \int^{\infty}_{a}  \frac{da}{H a^2}
\end{equation}
However, this definition of HDE can be modified to take account of quantum considerations [40]. Tsallis and Cirto have shown that the horizon entropy of a black hole may be modified as [27]:
\begin{equation}\label{eq:2.8}
S_{\delta} = \gamma A^{\delta} 
\end{equation}
where $\gamma$ is an unknown constant and $\delta$ denotes the non-additivity parameter. If we take $\gamma = 1/ 4G$ and $\delta = 1$, then the Bekenstein entropy can be recovered (where $\hbar = c = k_B =1$).\\
The holographic principle states that the number of degrees of freedom of a physical system should firstly scale with its bounding area rather than with its volume and secondly should be constrained by an infrared cutoff [21,22]. In ref.[23], Cohen et al. proposed that the inequality between the system entropy (S) and the IR (L) and UV ($\Lambda$) cutoffs should be defined as
\begin{equation}\label{eq:2.9}
L^3 \Lambda^3 \leq S^{3/4}
\end{equation} 
Combining with eq.$(2.8)$, we then have
\begin{equation}\label{eq:2.10}
\Lambda^4 \leq \gamma (4\pi)^{\delta} L^{2\delta - 4}
\end{equation}
where $\Lambda^4$ is the vacuum energy density. Based on the above inequality, the Tsallis holographic dark energy density (THDE) can be derived to be [30]
\begin{equation}\label{eq:2.11}
\rho_D = B L^{2\delta - 4}
\end{equation} 
where $B$ is an unknown parameter. If $B = 3 c^2 M^2_p$ and $\delta = 1$, then the energy density of holographic dark energy can be recovered.\\
We define the critical energy density $\rho_{cr}$ and the curvature energy density $\rho_k$ as usual as:
\begin{equation}\label{eq:2.12}
\rho_{cr}= 3 M^2_p H^2
\end{equation}
\begin{equation}\label{eq:2.13}
\rho_k = \frac{3k} {8\pi G a^2}
\end{equation}
We also introduce three fractional energy densities $\Omega_m$, $\Omega_D$ and $\Omega_k$:
\begin{equation}\label{eq:2.14}
\Omega_m = \frac{\rho_m} {\rho_{cr}} = \frac{\rho_m} {3 M^2_p H^2}
\end{equation} 
\begin{equation}\label{eq:2.15}
\Omega_D = \frac{\rho_D} {\rho_{cr}} = \frac{B} {3 M^2_p H^2} L^{2\delta -4}
\end{equation}
\begin{equation}\label{eq:2.16}
\Omega_k = \frac{\rho_k} {\rho_{cr}} = \frac{k} {H^2 a^2}
\end{equation}
Considering $L = H^{-1}$, we have
\begin{equation}\label{eq:2.17}
\Omega_D =  \frac{B} {3 M^2_p } L^{2\delta -2} = \frac{B} {3 M^2_p } H^{-2\delta +2}
\end{equation}
Now using eqs. $(2.3)$, $(2.4)$ and $(2.5)$, we obtain
\begin{equation}\label{eq:2.18}
\dot L = HL+a \dot r(t) = 1- \frac{1}{\sqrt{|k|}} \cos n (\sqrt{|k|} R_h / a) 
\end{equation}
In particular,
\begin{equation}\label{eq:2.19}
\cos n (\sqrt{|k|} x)= 
\begin{cases}
\cos(x)  & \text{$k$=1}\\
1 & \text{$k$=0}\\
\cosh(x) & \text{$k$=-1}
\end{cases}       
\end{equation}.\\ 
In flat space, if there is no interaction between Tsallis holographic dark energy and matter, i.e.,
\begin{equation}\label{eq:2.20}
\dot \rho_D + 3H \rho_D (1+\omega_D ) =0 
\end{equation}
\begin{equation}\label{eq:2.21}
\dot \rho_m + 3H \rho_m =0 
\end{equation}
then the equation of state for the Tsallis holographic energy density can be obtained as [41]:
\begin{equation}\label{eq:2.22}
\omega_D = \frac{\delta -1}{(2-\delta)\Omega_D - 1}
\end{equation}
For $\delta < 1$, we have $2-\delta> 1$ meaning that there is a divergence in the behavior of $\omega_D$ occuring at the red-shift for which $\Omega_D=1/(2-\delta)$. Therefore, the $\delta < 1$ case is not suitable in our setup [30].\\
In flat space, if there is an interaction between the Tsallis holographic dark energy and matter, i.e.,
\begin{equation}\label{eq:2.23}
\dot \rho_D + 3H \rho_D (1+\omega_D ) = -Q 
\end{equation}
\begin{equation}\label{eq:2.24}
\dot \rho_m + 3H \rho_m = Q
\end{equation}
where $Q=3 b^2 H (\rho_m + \rho_D)$ is the interaction term [41], the Tsallis holographic energy equation of state is then [41]:
\begin{equation}\label{eq:2.25}
\omega_D = \frac{\delta -1 + b^2 / \Omega_D}{(2-\delta)\Omega_D - 1}
\end{equation}

\section{The tachyon field as Tsallis holographic dark energy in a flat FRW universe}
\subsection{The non-interacting case}
Let us consider a four-dimensional, spacially-flat FRW universe, so that the Friedmann equations are 
\begin{equation}\label{eq:3.1}
H^2 = \frac{8 \pi G}{3} \rho
\end{equation}
\begin{equation}\label{eq:3.2}
\frac{\ddot a}{a} = - \frac{4\pi G (\rho + 3P)}{3}
\end{equation}
where $\rho = \rho_T + \rho_{NR} + \rho_{R}$ is the energy density for, tachyon matter, non-relativistic and relativistic matter, respectively, and $P$ is the corresponding pressure. In this subsection we shall restrict ourselves to considering a description of the present situation where we assume that tachyon field largely dominates the universe and therefore the energy density and pressure of non-relativistic and relativistic matter can be disregarded. Therefore, the first Friedmann equation eq.$(3.1)$ is then
\begin{equation}\label{eq:3.3}
H^2 = \frac{8 \pi G}{3} \rho_T
\end{equation} 
The energy density and pressure for the tachyon field are given by the following relations [25]:
\begin{equation}\label{eq:3.4}
\rho_T = \frac{V(T)}{\sqrt{1-\dot{T}^2}}
\end{equation}
\begin{equation}\label{eq:3.5}
P_T = -V(T) \sqrt{1-\dot{T}^2}
\end{equation}
where $V (T)$ is the potential energy of tachyon field. Combining eq.$(3.4)$ and $(3.5)$, we can obtain the equation of state:
\begin{equation}\label{eq:3.6}
\omega_T = \dot{T}^2 - 1
\end{equation}
We now propose a correspondence between the tachyon dark energy model and the Tsallis holographic dark energy scenario. In a flat universe, the density of Tsallis holographic dark energy is
\begin{equation}\label{eq:3.7}
\rho_D = B R^{2\delta - 4}_h
\end{equation}
where $R_h$ is given by eq.$(2.6)$ and 
\begin{equation}\label{eq:3.8}
\omega_D = \frac{\delta -1}{(2-\delta)\Omega_D - 1}
\end{equation}
which is given by eq.$(2.20)$.\\
If we establish a correspondence between the Tsallis holographic dark energy and the tachyon energy density, then using eq.$(3.4)$ and $(3.7)$, we have,
\begin{equation}\label{eq:3.9}
\rho_D = B R^{2\delta - 4}_h= \frac{V(T)}{\sqrt{1-\dot{T}^2}}
\end{equation}
Also using eq.$(3.6)$ and $(3.8)$, we can write
\begin{equation}\label{eq:3.10}
\omega_D =  \frac{\delta -1}{(2-\delta)\Omega_D - 1}=  \dot{T}^2 - 1
\end{equation}
so that
\begin{equation}\label{eq:3.11}
\dot{T} = \sqrt{\frac{(2-\delta) \Omega_D + \delta -2}{(2-\delta) \Omega_D -1}}
\end{equation}
Then we can reconstruct the function $V(T)$ and the variable $P_D$
\begin{equation}\label{eq:3.12}
V(T) = \rho_T \sqrt{1-\dot{T}^2} = B R^{2\delta - 4}_h \sqrt{\frac{1-\delta}{(2-\delta)\Omega_D -1 }}
\end{equation}
\begin{equation}\label{eq:3.13}
P_D = - V(T) \sqrt{1-\dot{T}^2} = - B R^{2\delta - 4}_h \frac{1-\delta}{(2-\delta)\Omega_D -1 }
\end{equation}
If there is no interaction between holographic dark energy and matter, we have
\begin{equation}\label{eq:3.14}
\dot \rho_D + 3H \rho_D (1+\omega_D ) =0 
\end{equation}
Using eq.$(2.17)$, in flat space, we have 
\begin{equation}\label{eq:3.15}
\dot L = 1- 1 = 0 
\end{equation}
Then we obtain that
\begin{equation}\label{eq:3.16}
\dot{\rho_D} = B (2\delta-4)L^{2\delta-5} \dot{L} =0 
\end{equation}
Inserting this result in eq.$(3.14)$, we have $\omega_D = -1$. Therefore, $\dot{T}^2 = 0$. \\
Inserting $\dot{T}^2 = 0$ into eq.$(3.11)$, we obtain that 
\begin{equation}\label{eq:3.17}
(2-\delta) \Omega_D + \delta -2 = 0
\end{equation} 
For which two possible solutions exist. One is $\delta = 2$. The other one is $\Omega_D = 1$, if $\delta \ne 2$.\\
If $\delta = 2$, based on eq.$(3.7)$, the energy density of dark energy is 
\begin{equation}\label{eq:3.18}
\rho_D = B
\end{equation} 
which is a constant. Then according to eq.$(3.12)$ and $(3.13)$, we have
\begin{equation}\label{eq:3.19}
 V(T) = B
\end{equation}
\begin{equation}\label{eq:3.20}
P_D = -B
\end{equation}
In this case $B$ is the value of cosmological constant $\Lambda$, which is commonly regarded as vacuum energy [42].\\
If $\Omega_D = 1$ and $\delta \ne 2$, then we obtain that
\begin{equation}\label{eq:3.21}
V(T) = B R^{2\delta - 4}_h
\end{equation}
\begin{equation}\label{eq:3.22}
P_D = -B R^{2\delta - 4}_h
\end{equation}
In both of the above two scenarios,  we have $\omega_D = -1$, which is consistent with the result of the cosmological constant $\Lambda$. However, we prefer the latter case since the former case cannot recover the Bekenstein entropy and since we have assumed that tachyon field dominates the universe. 

\subsection{The interacting case}
In this subsection we consider the interacting case for the model. As we have pointed out in section 2, in flat space, if there exists an interaction between the Tsallis dark energy and matter, we have
\begin{equation}\label{eq:3.23}
\dot \rho_D + 3H \rho_D (1+\omega_D ) = -Q 
\end{equation}
\begin{equation}\label{eq:3.24}
\dot \rho_m + 3H \rho_m = Q
\end{equation}
where $Q=3 b^2 H (\rho_m + \rho_D)$ is the interaction term and $b^2$ is a coupling parameter [41]. \\
In the same manner as in the non-interacting case, we have
\begin{equation}\label{eq:3.25}
\dot {\rho}_D = 0
\end{equation}
Then combining eq.$(3.23)$ and $Q=3 b^2 H (\rho_m + \rho_D)$, we derive that
\begin{equation}\label{eq:3.26}
\omega_D = -\frac{b^2}{\Omega_D} -1 
\end{equation}
Equating eq.$(3.26)$ with eq.$(3.6)$, we derive that
\begin{equation}\label{eq:3.27}
\dot{T}^2 = -\frac{b^2}{\Omega_D}  
\end{equation}
Therefore, if we require $\dot{T}^2$ to be real, then $b^2$ must be zero in flat space. Therefore, we can conclude that in flat space, in a tachyon model of Tsallis holographic dark energy, irrespective of whether or not there exists an interaction between dark energy and matter, $\dot{T}^2$ must always be zero. In addition, we can derive that
\begin{equation}\label{eq:3.28}
V(T) =\rho_D  
\end{equation}
\begin{equation}\label{eq:3.29}
P_D = -\rho_D  
\end{equation}
and then we conclude that in flat space, the equation of state $\omega_D$ is always $1$.

\section{The tachyon field as Tsallis holographic dark energy in a non-flat FRW universe}
In this section we extend the calculations of the previous section to the non-flat FRW universe.  In this case, the first Friedmann equation is given by
\begin{equation}\label{eq:4.1}
H^2 + \frac{k}{a^2} = \frac{1}{3 M^2_p} (\rho_m + \rho_D)
\end{equation}
where $k$ denotes the curvature of space so that $k= -1, 0, 1$ for closed, flat and open universe respectively.\\
Combining eq.$(2.17)$ and $(3.7)$, we obtain that
\begin{equation}\label{eq:4.2}
\dot{\rho_D} = B (2\delta-4)L^{2\delta-5} \dot{L} = B (2\delta-4)L^{2\delta-5} (1- \frac{1}{\sqrt{|k|}} \cos n (\sqrt{|k|} x) )
\end{equation}
where $x = R_h/a(t)$. Inserting eq.$(3.23)$, we obtain that
\begin{equation}\label{eq:4.3}
1- \dot{T}^2 = -\frac{1}{3} + b^2(1+u) + \frac{2}{3} \delta  - \frac{2\delta}{3}  \frac{1}{\sqrt{|k|}} \cos n (\sqrt{|k|} x) +\frac{4}{3} \frac{1}{\sqrt{|k|}} \cos n (\sqrt{|k|} x)
\end{equation}
namely,
\begin{equation}\label{eq:4.4}
\dot{T}^2 = \frac{4}{3} - b^2(1+u) - \frac{2}{3} \delta  + \frac{2\delta}{3}  \frac{1}{\sqrt{|k|}} \cos n (\sqrt{|k|} x) -\frac{4}{3} \frac{1}{\sqrt{|k|}} \cos n (\sqrt{|k|} x)
\end{equation}
where $1+u = \frac{1+\Omega_k}{\Omega_D}$ and $\Omega_D$ and $\Omega_k$ have been defined in eq.$(2.14)$ and $(2.15)$.\\
For the $k=1$ case, $\frac{1}{\sqrt{|k|}} \cos n (\sqrt{|k|} x) =\cos(x)$. Then eq.$(4.4)$ becomes 
\begin{equation}\label{eq:4.5}
\dot{T}^2 = \frac{4}{3} - b^2(1+u) - \frac{2}{3} \delta  + \frac{2\delta}{3} \cos(R_h/a) -\frac{4}{3} \cos(R_h/a)
\end{equation}
When $\dot{T}^2=0$, we have
\begin{equation}\label{eq:4.6}
\cos(R_h/a) = \frac{3b^2(1+u)}{2(\delta-2)} + \frac{2\delta -1}{2(\delta -2)}
\end{equation}
Eq.$(4.6)$ requires $0 \leq \cos(R_h/a) \leq 1$. In order to avoid divergence and recover the Bekenstein entropy, it is necessary to require that $1 \leq \delta<2$. However, according to eq.$(4.6)$, we have $\delta \leq \frac{1}{2}$, which is contradictory to our requirement.\\
Similarly, we can obtain the above equations for the $k = -1$ case whereby $\cos(x)$ should be replaced by $\cosh(x)$, namely,
\begin{equation}\label{eq:4.7}
\dot{T}^2 = \frac{4}{3} - b^2(1+u) - \frac{2}{3} \delta  + \frac{2\delta}{3} \cosh(R_h/a) -\frac{4}{3} \cosh(R_h/a)
\end{equation}
In particular, for the $k=-1$ case, when $\dot{T}^2=0$, 
\begin{equation}\label{eq:4.8}
\cosh(R_h/a) = \frac{3b^2(1+u)}{2(\delta-2)} + \frac{2\delta -1}{2(\delta -2)}
\end{equation} 
Since the value of $\cosh(x)$ is always larger than 1, we can derive that $b^2<0$, which is impossible. To conclude, in curved space, $\dot{T}^2$ cannot be zero, i.e., $\omega_D$ cannot be $-1$. In other words, this model cannot explain the origin of the cosmological constant $\Lambda$.\\
Figures 1, 2, 3 and 4 below are the evolution trajectories of $\dot{T}^2$ for different values of $\delta$ and $b^2$ for an open and a closed FRW universe, respectively. From figure 1, we observed that for an open universe, with the increasing of $\cos(R_h/a)$, $\dot{T}^2$s always decreases monotonically for different $\delta$s and are always larger than zero. While from figure 2, we know that for closed universe, with the increasing of $\cosh(R_h/a)$, $\dot{T}^2$ always decreases monotonically for the different values of $\delta$s and is always smaller than zero which is not a physically valid situation. From figures 3 and 4, we observe that with the increasing of $\cos(R_h/a)$ and $\cosh(R_h/a)$, $\dot{T}^2$ always decreases monotonically for the different values of $b^2$.  
\begin{figure}[htbp]
	\centering
	\includegraphics[scale=0.4]{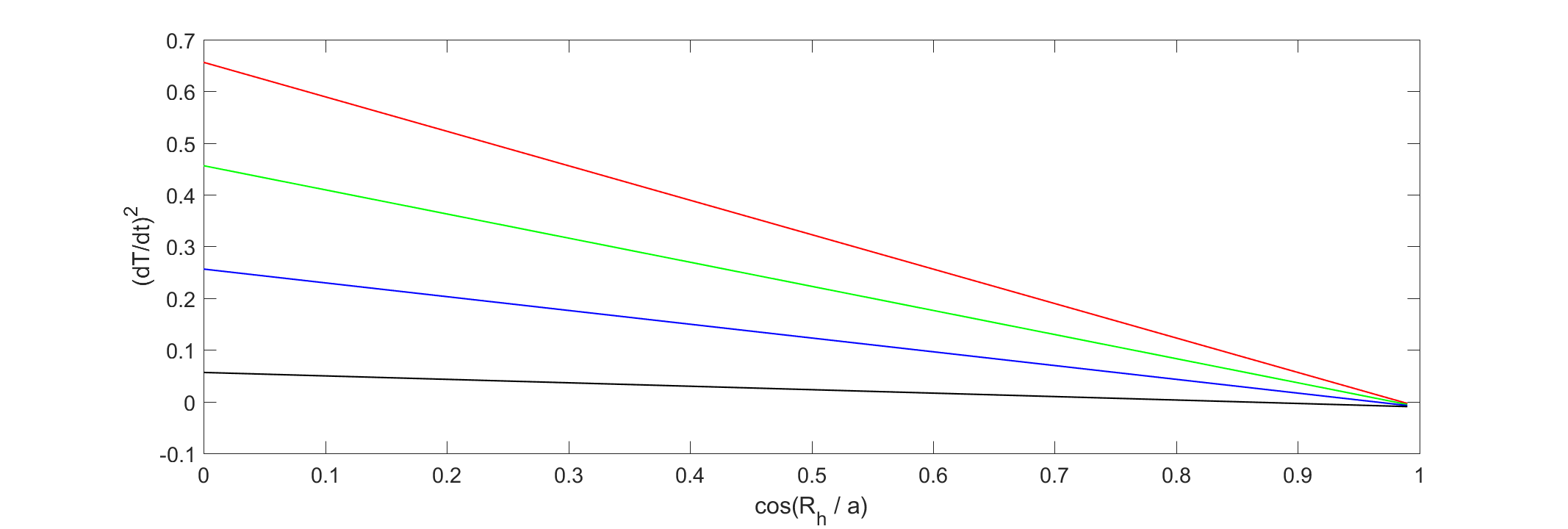}
	\caption{Evolution trajectories of $\dot{T}^2$ for different values of $\delta$ for an open FRW universe. Here we set $u=0.04$. The red line corresponds to $\delta = 1$, the green line to $\delta = 1.3$, the blue line to $\delta = 1.6$ and the black line to $\delta = 1.9$.}
\end{figure}\\
\begin{figure}[htbp]
	\centering
	\includegraphics[scale=0.4]{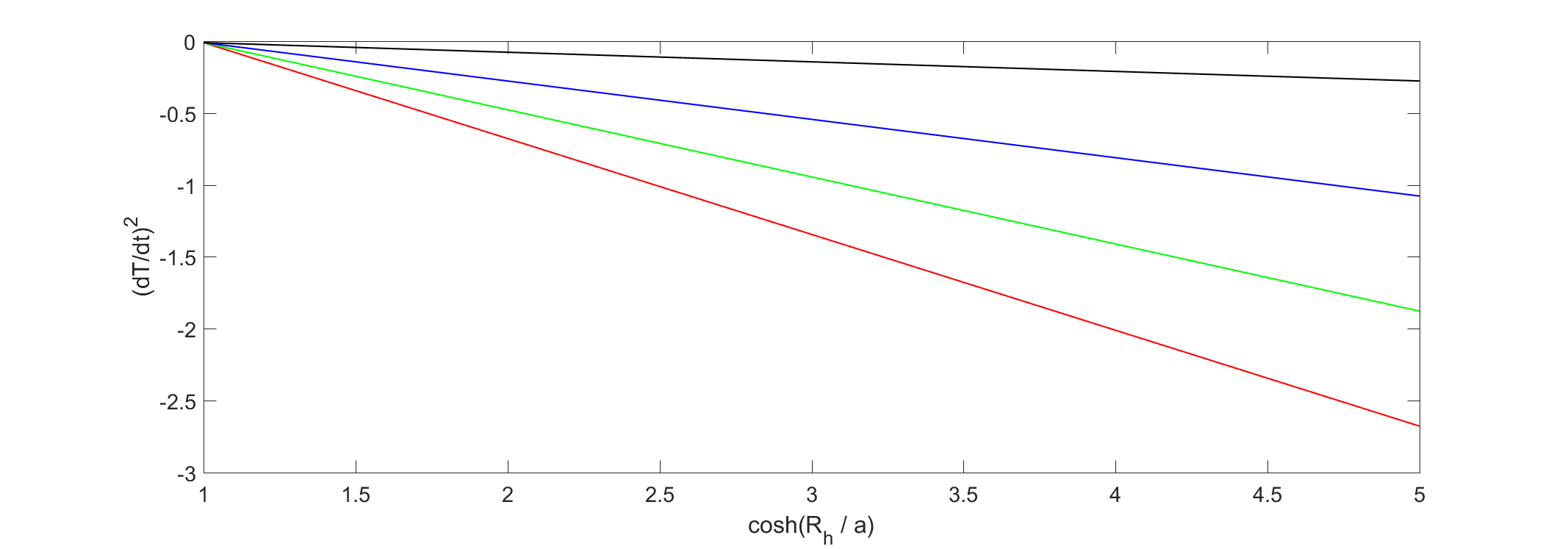}
	\caption{Evolution trajectories of $\dot{T}^2$ for different values of $\delta$ for a closed FRW universe. Here we set $u=0.04$. The red line corresponds to $\delta = 1$, the green line to $\delta = 1.3$, the blue line to $\delta = 1.6$ and the black line to $\delta = 1.9$.}
\end{figure}\\
\begin{figure}[htbp]
	\centering
	\includegraphics[scale=0.4]{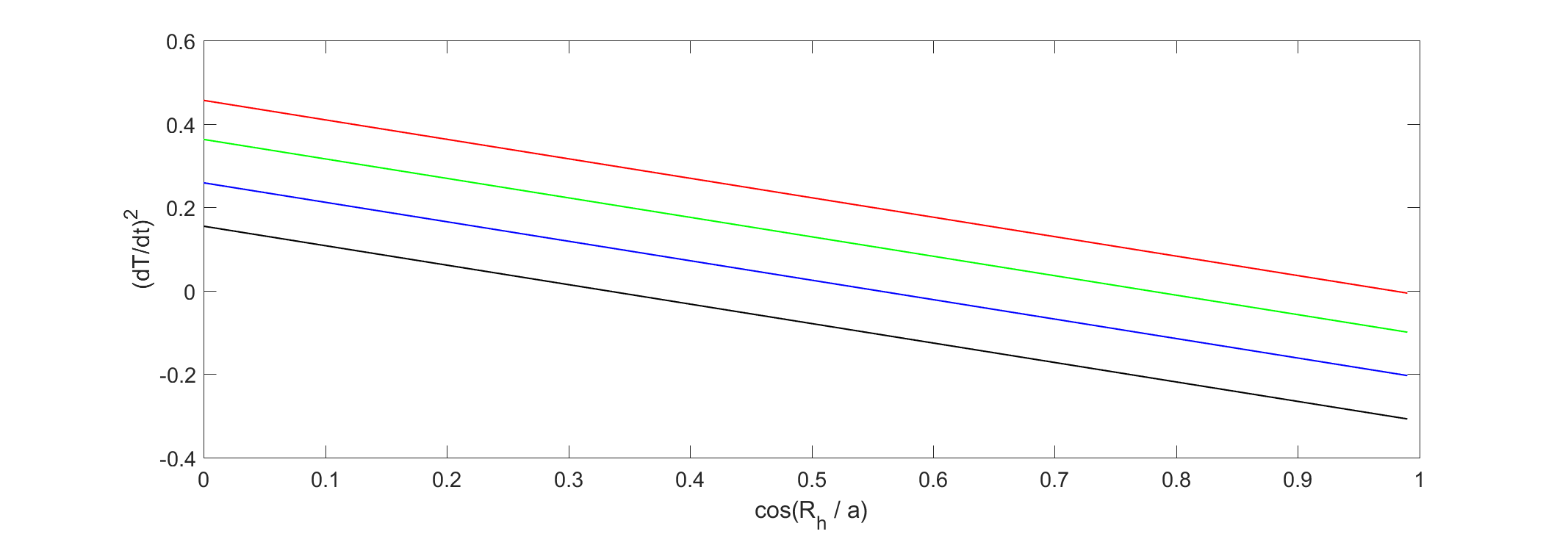}
	\caption{Evolution trajectories of $\dot{T}^2$ for different values of $b^2$ for an open FRW universe. Here we set $u=0.04$. The red line corresponds to $b^2 = 0.01$, the green line to $b^2 = 0.10$, the blue line to $b^2= 0.20$ and the black line to $b^2 = 0.30$.}
\end{figure}\\
\begin{figure}[htbp]
	\centering
	\includegraphics[scale=0.4]{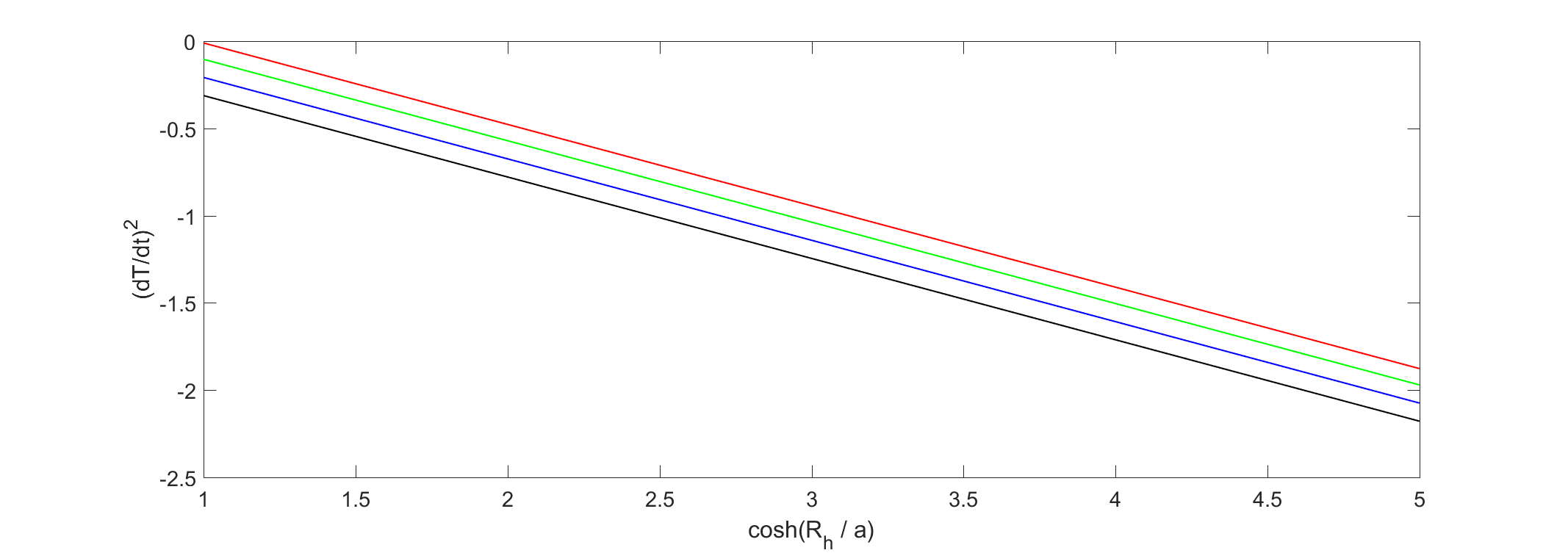}
	\caption{Evolution trajectories of $\dot{T}^2$ for different values of $b^2$ for a closed FRW universe. Here we set $u=0.04$. The red line corresponds to $b^2 = 0.01$, the green line to $b^2 = 0.10$, the blue line to $b^2= 0.20$ and the black line to $b^2 = 0.30$.}
\end{figure}\\

\section{Summary and Discussion} 
To address the problem of accounting for the accelerating expansion of the universe and due to an absence of knowledge in this domain, theoretical cosmologists have considered a variety of dark matter candidates to explain this phenomenon. The holographic dark matter (HDE) model which was proposed by Fischler and Susskind [43] has been widely studied [44]. In recent years one of the main research directions has involved the use of the Tsallis holographic dark energy model [30,45,46,47]. At first glance, it appears to be an appropriated  model for the current universe in the standard cosmology framework [30,32,33]. However, in the same way as the primary HDE based on the Bekenstein entropy is unstable (see [34]), THDE is also unstable [30,32,33]. More studies on the various cosmological features of the Tsallis generalized statistical mechanics can be found in ref.[35]. Most past papers considered the effects of HDE in different real scalar field theory models, such as quintessence [9], K-essence [10], tachyon [12] etc. For example, in ref.[25], the authors considered the possibility of using the tachyon model as a holographic dark energy model. In the present article, we consider the further possibility of the tachyon field model as a Tsallis holographic dark energy model. Most past papers considered the effects of HDE in different real scalar field theory, such as quintessence [9], K-essence [10], tachyon [12] etc. For example, in ref.[25], the authors considered the possibility of tachyon model as holographic dark energy model. In this article, we further consider the possibility of tachyon field model as Tsallis holographic dark energy model. \\ 
We have proposed a correspondence between the Tsallis holographic dark energy scenario and the tachyon field model in a flat and in a non-flat FRW universe, respectively. We then reconstructed the potential $V (T)$ and considered the dynamics of the tachyon field which describes cosmology. We find that in the case of a flat universe, in a tachyon model of Tsallis holographic
dark energy, irrespective of whether there exists an interaction between dark energy and matter or not, $\dot{T}^2$ must always be zero. Therefore, the equation of state $\omega_D$ is always $-1$ in flat universe. In the case of a non-flat universe,  $\dot{T}^2$ cannot be zero so that $\omega_D \neq -1$, which cannot be used to explain the origin of the cosmological constant. $\dot{T}^2$ monotonically decreases with the increasing of $\cos(R_h/a)$ and $\cosh(R_h/a)$ for different values of $\delta$. In particular, for an open universe, $\dot{T}^2$ is always larger than zero whereas for a closed universe, $\dot{T}^2$ is always smaller than zero which is not a physically valid situation. In addition, we conclude that, with the increasing of $\cos(R_h/a)$ and $\cosh(R_h/a)$, $\dot{T}^2$ always decreases monotonically for different $b^2$s.\\
Future work can develop this research along the following directions. Firstly, we can establish the correspondence between the tachyon field and other dark energy scenarios, in particular for a stable dark energy model and compare the results with observational data. Secondly, we can expand  the possibility of using a real scalar field as the dark energy model considered in this article to  explore the possibilities of using complex scalar fields as the dark energy model. In fact, in ref.[48], we have already considered the possibility of ghost dark energy as a complex quintessence field. We intend to explore further complex scalar field possibilities as dark energy candidates and form a framework of these candidates. Finally, we also need to consider how to ensure compatibility of these results with fundamental theories, such as string theory and loop quantum gravity. There is therefore great potential for development of this work in the future.\\







\end{document}